\documentclass[runningheads,envcountreset,envcountsect,10pt]{llncs}

\newcounter{myquestion}

\usepackage{url}
\usepackage{color}
\usepackage[font=small]{caption}
\usepackage{subfigure}
\usepackage[pdftex]{graphicx}
\usepackage[cmex10]{amsmath}
\usepackage{amssymb}
\usepackage{cite}
\usepackage{epstopdf}

\newcommand{\proba}{\mathbf{P}}
\newcommand{\Prob}[3]{\mathit{Prob}^{#1}_{#2(#3)}}
\newcommand{\Probs}[2]{\mathit{Prob}^{#1}_{#2}}
\newcommand{\probmeasure}{\mathit{\Pr}}
\newcommand{\filter}{\mathit{filter}}

\newcommand{\feed}{\mathit{feed}}
\newcommand{\pick}{\mathit{pick}}
\newcommand{\seeY}{\mathit{seeY}}
\newcommand{\seeP}{\mathit{seeP}}

\newcommand{\ra}{\rightarrow}
\newcommand{\lra}{\longrightarrow}

\newcommand{\AP}{\mathit{AP}}

\newcommand{\Path}{\mathit{Path}}
\newcommand{\D}{\mathcal{D}}

\newcommand{\init}{\mathit{init}}

\newcommand{\initdist}{\iota^{\mathit{init}}}

\newcommand{\umm}{\mathcal{M}}

\newcommand{\TF}{{\sf F}}
\newcommand{\TU}{{\sf U}}

\newcommand{\TX}{{\sf X}}

\newcommand{\PRISM}{{\sf PRISM}}

\ifx\definition\undefined
\newtheorem{definition}{Definition}
\fi

\ifx\example\undefined
\newtheorem{example}{Example}
\fi

\ifx\proposition\undefined
\newtheorem{proposition}{Proposition}
\fi

\begin{document}
\pagestyle{headings}

\mainmatter 

\title{Probabilistic Model Checking of DTMC Models of User Activity
  Patterns}

\titlerunning{Probabilistic Model Checking of DTMC Models of User
  Activity Patterns}

\author{Oana Andrei\inst{1}\and Muffy Calder\inst{1} \and Matthew
  Higgs\inst{1} \and Mark Girolami\inst{2}} \authorrunning{Andrei et
  al.}

\institute{School of Computing Science, University of Glasgow, G12
  8RZ, UK \and Department of Statistics, University of Warwick,
   CV4 7AL, UK}

\maketitle

\begin{abstract}
  Software developers cannot always anticipate how users will actually
  use their software as it may vary from user to user, and even from
  use to use for an individual user. In order to address questions
  raised by system developers and evaluators about software usage, we
  define new probabilistic models that characterise user
  behaviour, based on activity patterns inferred from actual logged
  user traces.  We encode these new models in a probabilistic model
  checker and use probabilistic temporal logics to gain insight into
  software usage. We motivate and illustrate our approach by
  application to the logged user traces of an iOS app.
\end{abstract}

\section{Introduction}

Software developers cannot always anticipate how users will {\em
  actually} use their software, which is sometimes surprising and
varies from user to user, and even from use to use, for an individual
user. We propose that temporal logic reasoning over formal,
probabilistic models of actual logged user traces can aid software
developers, evaluators, and users by:
\begin{itemize}
\item providing insights into application usage, including differences
  and similarities between different users,
\item predicting   user
behaviours, and 
\item recommending future application development.
\end{itemize}

Our approach is based on systematic and automated logging and
reasoning about users of applications. While this paper is focused on
mobile applications (apps), much of our work applies to any software
system. A logged user trace is a chronological sequence of in-app
actions representing how the user explores the app. From logged user
traces of a population of users we infer {\em activity patterns},
represented each by a discrete-time Markov chain (DTMC), and for each
user we infer a user {\em strategy} over the activity patterns.  For
each user we deduce a meta model based on the set of all activity
patterns inferred from the population of users and the user strategy,
and we call it the {\em user metamodel}. We reason about the user
metamodel using probabilistic temporal logic properties to express
hypotheses about user behaviours and relationships within and between
the activity patterns, and to formulate {\em app-specific} questions
posed by developers and evaluators.

We motivate and illustrate our approach by application to the mobile,
multiplayer game Hungry Yoshi~\cite{McMillanMBHC10}, which was
deployed in 2009 for iPhone devices and has involved thousands of
users worldwide.  We collaborate with the Hungry Yoshi developers on
several mobile apps and we have access to all logged user data. We
have chosen the Hungry Yoshi app because its functionality is
relatively simple, yet adequate to illustrate how formal analysis can
inform app evaluation and development.

The main contributions of the paper are:
\begin{itemize}
\item a formal and systematic approach to formal user activity
  analysis in a probabilistic setting,
\item inference of user activity patterns represented as DTMCs,
\item definition of the DTMC user metamodel and guidelines for
  inferring user metamodels from logged user data,
\item encoding of the user metamodel in the \PRISM~model checker and
  temporal logic properties defined over both states and activity
  patterns as atomic propositions,
\item illustration with a case study of a deployed app with thousands
  of users and analysis results that reveal insights into real-life
  app usage.
\end{itemize}

The paper is organised as follows.  In the next section we give an
overview of the Hungry Yoshi app, which we use to motivate and
illustrate our work.  We list some example questions that have been
posed by the Yoshi developers and evaluators; while these are specific
to the Yoshi app, they are also indicative questions for any app.  In
Sect.~\ref{sect:background} we give background technical definitions
concerning DTMCs and probabilistic temporal logics.  In
Sect.~\ref{inference} we define inference of user activity patterns,
giving a small example as illustration and some example results for
Hungry Yoshi.  In Sect.~\ref{umm} we define the user metamodel, we
illustrate it for Hungry Yoshi and we give an encoding for the
\PRISM~model checker.  In Sect.~\ref{analysis} we consider how to
encode some of the questions posed in Sect.~\ref{questions} in
probabilistic temporal logic, and give some results for an example
Hungry Yoshi user metamodel.  In Sect.~\ref{discussion} we reflect
upon the results obtained for Hungry Yoshi and some further issues
raised by our approach.  In Sect.~\ref{related} we review related work
and we conclude in Sect.~\ref{conclusions}.

\section{Running example:  Hungry Yoshi}\label{yoshi}

\begin{figure*}[!t]
  \centering 
  \subfigure[Main menu]{\includegraphics[scale=0.16]{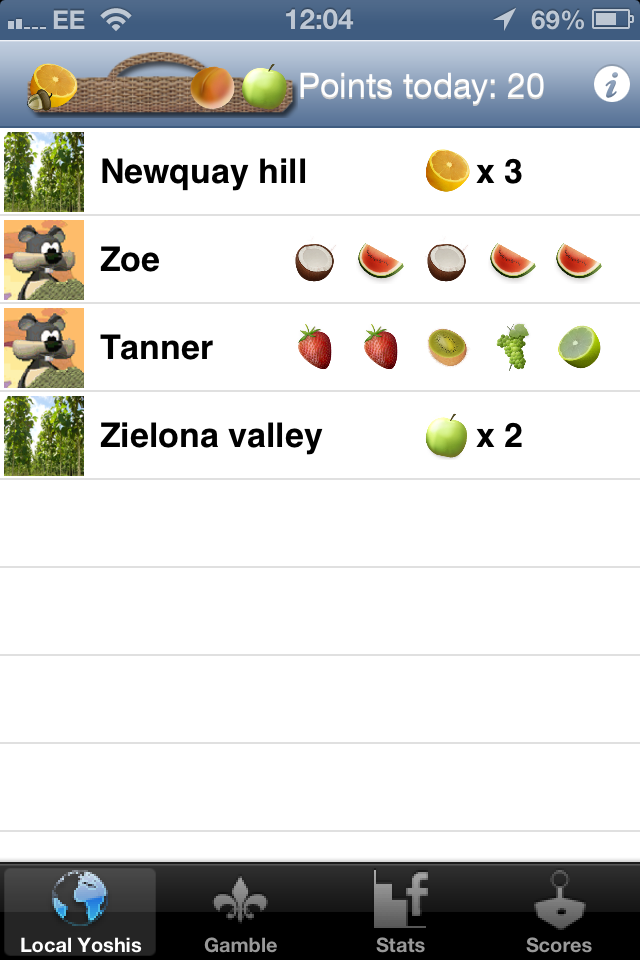}}
  \quad
  \subfigure[Yoshi Zoe]{\includegraphics[scale=0.16]{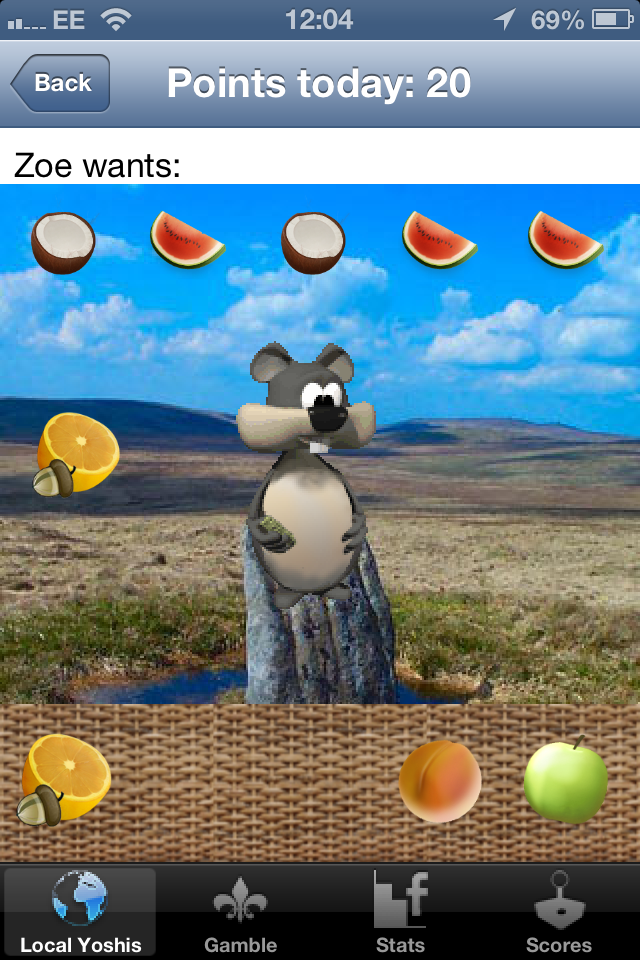}}
  \quad
  \subfigure[Newquay plantation]{\includegraphics[scale=0.16]{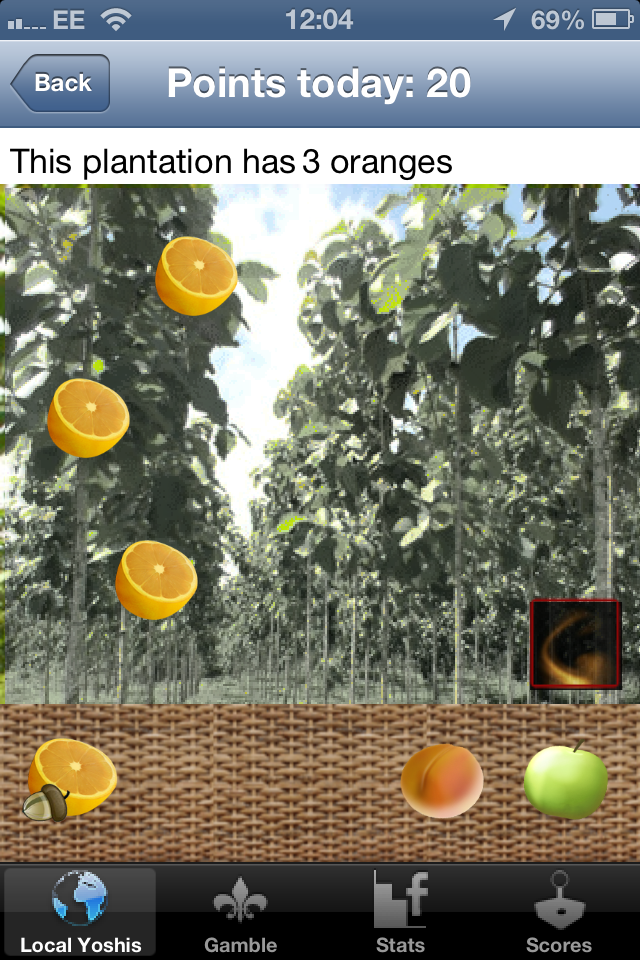}} 
  \caption{Hungry Yoshi screenshots: two plantations (Newquay hill and
    Zielona valley) and two yoshis (Zoe and Taner) are observed. The
    main menu shows the available plantations and yoshis with their
    respective content and required types of fruit. The current basket
    contains one orange seed, one apricot and one apple.}
\label{fig:screenshots}
\end{figure*}

The mobile, multiplayer game Hungry Yoshi~\cite{McMillanMBHC10} is
based on picking pieces of fruit and feeding them to creatures called
{\em yoshis}. Players' mobile devices regularly scan the available
WiFi access points and display a password-protected network as a {\em
  yoshi} and a non-protected network as a {\em fruit
  plantation}. Plantations grow particular types of {\em fruit}
(possibly from {\em seeds}) and yoshis ask players for particular
types of fruit. Players score points if they pick the fruit from the
correct plantations, store them in a basket, and give them to yoshis
as requested.  There is further functionality, but here we concentrate
on the key user-initiated events, or {\em button taps}, which are:
{\em see a yoshi}, {\em see a plantation}, {\em pick fruit} and {\em
  feed a yoshi}. The external environment (as scanned by device),
combined with user choice, determines when yoshis and plantations can
be observed. The game was instrumented by the developers using the
SGLog data logging infrastructure~\cite{Hall:2009}, which streams logs
of specific user system operations back to servers on the developing
site as user traces. The developers specify directly in the source
code what method calls or contextual information are to be logged by
SGLog. A sample of screenshots from the game is shown in
Fig.~\ref{fig:screenshots}.

\subsection{Example questions from developers and
  evaluators} \label{questions}

Key to our formal analysis is suitable hypotheses, or {\em questions},
about user behaviour.  For Hungry Yoshi, we interviewed the developers
and evaluators of the game to obtain questions that would provide
useful insights for them.  Interestingly most of their hypotheses were
{\em app-specific}, and so we focus on these here, and then indicate
how each could be generalised.  We note that to date, tools available
to the developers and evaluators for analysis include only SQL and
iPython stats scripts.
 
\begin{enumerate}

\item When a yoshi has been fed $n$ pieces of fruit (which results in
  extra points when $n$ is 5), did the user interleave {\em pick
    fruit} and {\em feed a yoshi} $n$ times or did the user perform $n$
  {\em pick fruit} events followed by $n$ {\em feed a yoshi} events?
  And afterwards, did he/she continue with that pick-feed strategy or
  change to another one?  Which strategy is more likely in which
  activity pattern?  More generally, when there are several ways to
  reach a goal state, does the user always take a particular route and
  is this dependent on the activity pattern? 
 
\item If a user in one activity pattern does not {\em feed a yoshi}
  within $n$ button taps, but then changes to another activity
  pattern,  is the user then likely to {\em feed a yoshi} within $m$
  button taps?  More generally, which events cause a change of
  activity pattern, and which events follow that change of activity
  pattern? 

\item What kind of user tries to {\em pick fruit} $6$ times in a row
  (a basket can only hold $5$ pieces of fruit)?  More generally, in
  which activity pattern is a user more likely to perform an
  inappropriate event? 

\item If a user reads the instructions once, then does that user  reach a goal
  state in fewer steps than a user who does not read the instructions
  at all?  (Thus indicating the instructions are of some utility.)
  More generally, if a user performs a given event, then it is more
  likely that he/she will perform another given event, within $n$ button
  taps, than users that have not performed the first event?  Is this
  affected by the activity pattern? 

\end{enumerate}

\section{Technical Background}
\label{sect:background}
We assume familiarity with Discrete-Time Markov Chains, probabilistic
logics PCTL and PCTL*, and model
checking~\cite{BaierKatoen-MCbook,KwiatkowskaNP07}; basic definitions
are below.

A {\em discrete-time Markov chain} (DTMC) is a tuple
$\D=(S,\bar{s},\proba,L)$ where: $S$ is a set of states; $\bar{s}\in
S$ is the initial state; $\proba:S\times S \ra [0,1]$ is the
transition probability function (or matrix) such that for all states
$s\in S$ we have $\sum_{s'\in S}\proba(s,s')=1$; and $L:S\ra 2^{AP}$
is a labelling function associating to each state $s$ in $S$ a set of
valid atomic propositions from a set $\AP$.  A {\em path} (or
execution) of a DTMC is a non-empty sequence $s_0s_1\ldots$ where
$s_i\in S$ and $\proba(s_i,s_{i+1})>0$ for all $i\geq 0$. A path can
be finite or infinite. Let $\Path^\D(s)$ denote the set of all
infinite paths of $\D$ starting in state $s$.

Probabilistic Computation Tree Logic (PCTL)~\cite{BaierKatoen-MCbook}
and its extension PCTL* allow one to express a probability measure of
the satisfaction of a temporal property.  Their syntax is the
following:
\begin{center}
  \begin{tabular}{rl} 
    {\em State formulae} & $\quad \Phi ::=
    \mathit{true} \mid a \mid \lnot \Phi \mid \Phi \land \Phi \mid
    \proba_{\bowtie\, p}[\Psi]$
    \\
    {\em PCTL Path formulae} & $\quad \Psi ::= \TX\, \Phi \mid \Phi\,
    \TU^{\leq n}\, \Phi$
    \\
    {\em PCTL* Path formulae} & $\quad \Psi ::= \Phi\ \mid \Psi \land
    \Psi \mid \lnot \Psi \mid \TX\, \Psi \mid \Psi \TU^{\leq n}\,
    \Psi$
\end{tabular}
\end{center}
where $a$ ranges over a set of atomic propositions $\mathit{AP}$,
$\bowtie\,\in \{\leq, <, \geq, >\}$, $p\in [0,1]$, and $n\in
\mathbb{N}\cup \{\infty\}$. 

A state $s$ in a DTMC $\D$ satisfies an atomic proposition $a$ if
$a\in L(s)$.  A state $s$ satisfies a state formula $\proba_{\bowtie\,
  p}[\Psi]$, written $s\models \proba_{\bowtie\, p}[\Psi]$, if the
probability of taking a path starting from $s$ and satisfying $\Psi$
meets the bound $\bowtie\, p$, i.e., $\probmeasure_s\{\omega\in
\Path^\D(s) \mid \omega\models \Psi\} \bowtie\, p$, where
$\probmeasure_s$ is the probability measure defined over paths from
state $s$.  The path formula $\TX\, \Phi$ is true on a path starting
with $s$ if $\Phi$ is satisfied in the state following $s$; $\Phi_1\,
\TU^{\leq n}\, \Phi_2$ is true on a path if $\Phi_2$ holds in the
state at some time step $i\leq n$ and at all preceding states $\Phi_1$
holds. This is a minimal set of operators, the propositional operators
$\mathit{false}$, disjunction and implication can be derived using
basic logical equivalences and a common derived path operators is the
\emph{eventually} operator $\TF$ where $\TF^{\leq n}\,\Phi\equiv
\mathit{true}\, \TU^{\leq n}\,\Phi$.  If $n =\infty$ then superscripts
omitted.
 
We assume the following two additional notations.  Let $\varphi$
denote the state formulae from the propositional logic fragment of
PCTL, i.e., $\varphi ::= \mathit{true} \mid a \mid \lnot \varphi \mid
\varphi \land \varphi$, where $a\in \AP$.  Let $\D_{\mid \varphi}$
denote the DTMC obtained from $\D$ by restricting the set of states to
those satisfying $\varphi$.

Many of the properties we will examine require PCTL*, because we want
to examine sequences of events: this requires multiple occurrences of
a bounded until operator.  This is not fully implemented in the
current version of \PRISM\ (only a single bounded $\TU$ is
permitted\footnote{Because currently the LTL-to-automata translator
  that PRISM uses does not support bounded properties.})  and so we
combine probabilities obtained from \PRISM\ {\em filtered properties}
to achieve the same result.  Filtered probabilities check for
properties that hold {\em from sets of states} satisfying given
propositions.  For a DTMC $\D$, we define the filtered probability of
taking all paths that start from any state satisfying $\varphi$ and
satisfy (PCTL) $\psi$ by:
$$
\Prob{\D}{\filter}{\varphi}(\psi)\stackrel{\mathit{def}}{=}
\filter_{s\in\D, s\models \varphi}\probmeasure_s\{\omega\in \Path^\D(s) \mid
\omega \models \psi\}
$$
where $\filter$ is an operator on the probabilities of $\psi$ for all
the states satisfying $\varphi$. In the examples illustrated in this
paper we always apply the function $\mathit{min}$ as the filter
operator. If $\varphi$ uniquely identifies a state, then the filters
$\mathit{min}$, $\mathit{max}$, $\mathit{avg}$ give the same result,
and so we omit the filter from formulae.

\section{Inferring User Activity Patterns}\label{inference}
The role of inference is to construct a representation of the data
that is amenable to checking probabilistic temporal logic
properties. Developers want to be able to select a user and explore
that user's model.  While this could be achieved by constructing an
independent DTMC for each user, there is much to be gained from
sharing information between users. One way to do this is to construct
a set of {\em user classes} based on attribute information, and to
learn a DTMC for each class. This is the approach taken
in~\cite{GhezziPST14} for users interacting with web applications, and
is a natural way to aggregate information over users and to condition
user-models on attribute values.  One issue with this approach is that
it assumes within-class use to be homogeneous.  For example, all users
in the same city using the same browser are modelled using the same
DTMC.

In this work we take a different approach to inference. We construct
user models based on the log information alone, without any ad-hoc
specification of user classes. One reason for doing this is that we
have found the common representations of context - such as location,
operating system, or time of day - to be poor predictors of mobile
application use.  We would like to be able to construct user classes
based on more sophisticated representations of context, such as a
user's level of development or engagement in the app, or their
tendency to particular activity patterns.  By letting the data speak
for itself, we hope to uncover interesting activity patterns and
meaningful representations of users.

\subsection{Probabilistic model and inference}

We extend the standard DTMC model by introducing a {\em strategy} for
each user over activity patterns.  More formally, we assume there
exists a finite $K$ number of activity patterns, each modelled by a
DTMC denoted $\alpha_{k}=(S,\iota^{init},\proba_{k},L)$, for $k =
1,\ldots,K$.  Note only the transition probability $\proba_{k}$ varies
over the set of DTMCs, all the other parameters are the same.  For
some enumeration of users $m=1,\ldots, M$, we represent a user's
strategy by a vector $\theta_{m}$ such that $\theta_{m}(k)$ denotes
the probability that user $m$ transitions according to the DTMC
indexed by $k$.  We assume all DTMCs are available to all users at all
points in time.

The data for each user is assumed to be generated in the following
iterative way.  An initial state is chosen according to
$\initdist$. When in state $s \in S$, user $m$ selects the $k$th DTMC
with probability $\theta_m(k)$.  If the user chooses the $k$th DTMC,
then they transition from state $s$ to $s' \in S$ with probability
$\proba_k(s, s')$. It is assumed that each user is independent from
all other users.  This simple description specifies all the
probabilistic dependencies in our statistical model.  The power of
this model is that it represents users as mixtures of activity
patterns, and enables us to examine users in the low dimensional space
of $\theta_{m}$.

While there are many ways to infer the parameters of this model, we
choose simply to find the parameters that make the data most
likely. This requires us to use the expectation-maximisation (EM)
algorithm of~\cite{Demp1977}. The EM algorithm involves optimising a
non-linear objective and, to remedy this, we restart the algorithm
multiple times and use the optimal parameters over all restarts.

\subsection{Example activity patterns from Hungry Yoshi}

\begin{figure}[!ht]
  \centering
\includegraphics[height=45mm]{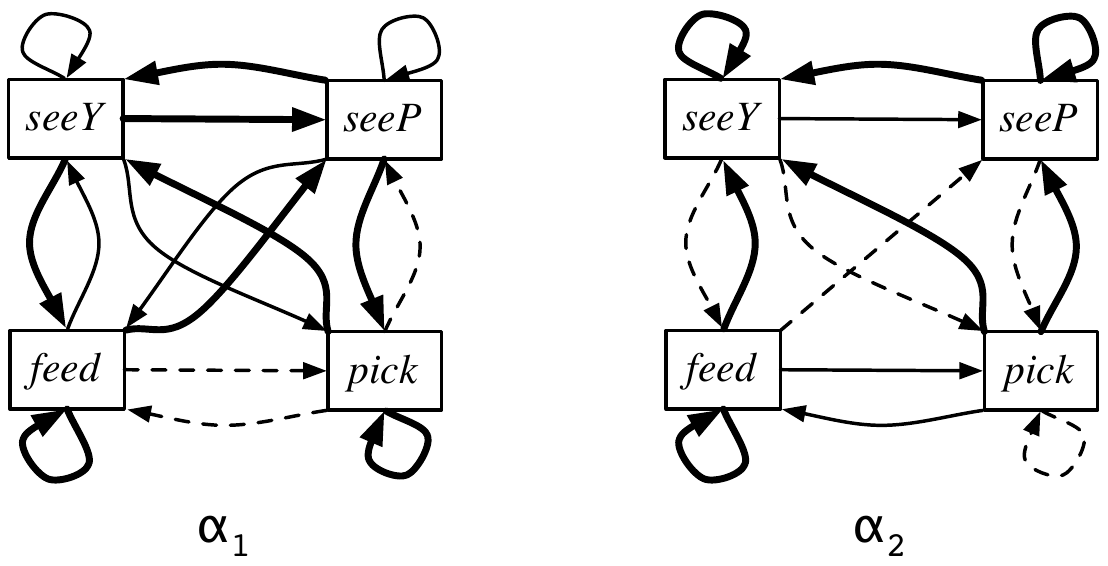}
\caption{Two user activity patterns $\alpha_1$ and $\alpha_2$ inferred
  from Hungry Yoshi.}
\label{fig:yoshi-patterns}
\end{figure}

In Fig.~\ref{fig:yoshi-patterns} we give the activity patterns,
inferred from a dataset of user traces for 164 users randomly selected
from the user population, for $K=2$. A more detailed overview is given
in the work-in-progress paper~\cite{HiggsMGC13}. For brevity, we do
not include the exact values of $\proba_1$ and $\proba_2$, but thicker
arcs correspond to transition probabilities greater than $0.1$,
thinner ones to transition probabilities in $[0.01, 0.1]$, and dashed
ones to transition probabilities smaller than $10^{-12}$.
Intuitively, we can see that given the game is essentially about
seeing yoshis and feeding them, $\alpha_{1}$ looks like a better way 
for playing the game. For example in $\alpha_{2}$ it is quite rare to
reach {\em feed} from {\em seeY} and {\em seeP}, and also rare to move
from {\em seeP} to {\em pick}.  Hungry Yoshi is a simple app with only
two distinctive activity patterns, in a more complex setting we might
not be able to have any intuition about the activity patterns.

\section{User Metamodel}\label{umm}

We define the formal model of the behaviour of a user $m$ with respect
to the population of users, which we call the {\em user metamodel}
(UMM).  The UMM for user $m$ is a DTMC where at each time step the
user either follows the current activity pattern $\alpha_k$ with
probability $\theta_m(k)$ and makes a move according to $\alpha_k$
{\em or} selects a different activity pattern $\alpha_{k'}$ with
probability $\theta_m(k')$ and then makes a move according to
$\alpha_{k'}$. Basically, we compose the $K$ DTMCs modelling the user
activity patterns using the strategy $\theta_m$.

\begin{definition}[User Metamodel]
  Given $K$ activity patterns $\alpha_1,\ldots,\alpha_K$ and
  $\theta_m$ the strategy of user $m$ for choosing activity patterns,
  the {\em user metamodel} for $m$ is a DTMC
  $\umm=(S_\umm,\iota^{init}_\umm,\proba_\umm,\mathcal{L}_\umm)$
  where:
\begin{itemize}
\item $S_\umm=S\times \{1,\ldots,K\}$,
\item $\iota^{\mathit{init}}_\umm(s,k) = \theta_m(k) \cdot
  \iota^{\mathit{init}}(s)$,
\item $\proba_\umm((s,k),(s',k'))=
  \theta_m(k')\cdot\proba_{k'}(s,s')$,
\item $\mathcal{L}_\umm(s,k) = \mathcal{L}(s) \cup \{\alpha=k\}$.
\end{itemize}
\end{definition}

\noindent 
We label each state $(s,k)$ with the atomic proposition $\alpha=k$ to
denote that the state belongs to the activity pattern $\alpha_k$.

\subsection{Example UMM from Hungry Yoshi}

\begin{figure}[!ht]
  \centering
\includegraphics[height=50mm]{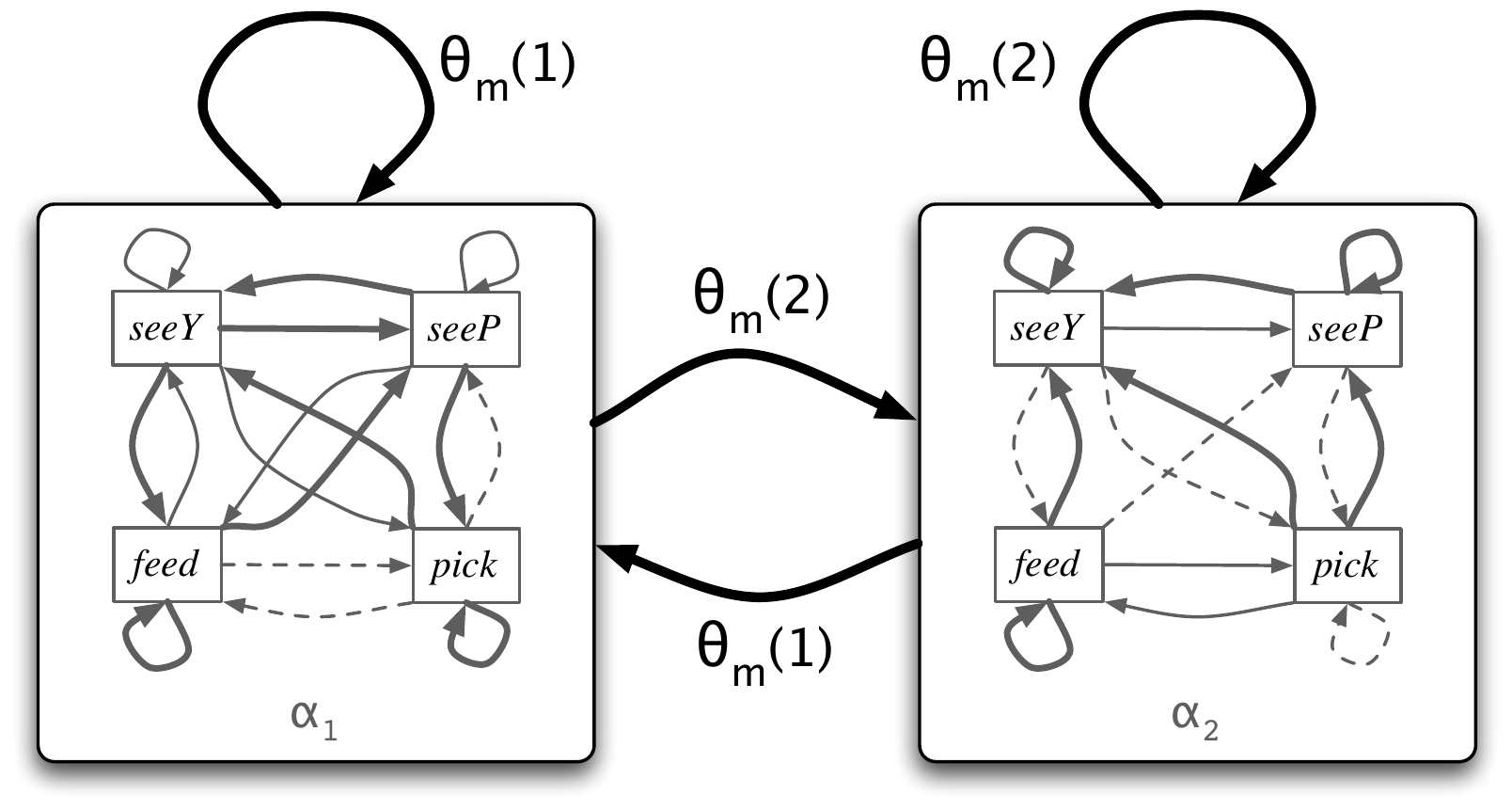}
\caption{An intuitive view of computing the transition probability
  matrix of the user metamodel for the Hungry Yoshi game.}
\label{fig:yoshi-umm}
\end{figure}

An intuitive graphical description of the UMM for the Hungry Yoshi
game for $K=2$ is illustrated in Fig.~\ref{fig:yoshi-umm}. For
example, $\theta_m(1)$ is the probability that user $m$ continues with
activity pattern $\alpha_{1}$, i.e. takes a transition between states
in $\alpha_{1}$.  The probability that the user changes the activity
pattern and makes a transition according to $\alpha_{2}$ is
proportional to $\theta_m(2)$. Figure~\ref{fig:yoshi-umm} is not a
direct representation of the transition probability matrix of the UMM
DTMC, but it illustrates how that matrix is derived from the matrices
of the individual user activity patterns. Note that the activity
patterns have the same sets of states.  For instance, in the Hungry
Yoshi example, consider we are in state $\seeY$ with $\alpha_{1}$; we
can move to state $\feed$ following the same pattern $\alpha_{1}$ with
the probability $\theta_{m}(1) \cdot P_{1}(\seeY,\feed)$, \emph{or} we
can change the activity pattern and move to state $\feed$ following
$\alpha_{2}$ with the probability $\theta_{m}(2) \cdot
P_{2}(\seeY,\feed)$.

\subsection{Encoding a UMM in   \PRISM}\label{encoding}
We use the probabilistic model checker \PRISM~\cite{PRISM}.  We assume
some familiarity with the modelling language (based on the language of
reactive modules), which includes global variables, modules with local
variables, labelled-commands corresponding to transitions and multiway
synchronisation of modules. In Fig.~\ref{fig:prism-encoding} we
illustrate the PRISM encoding of the UMM for user $m$. The
representation is straightforward, consisting of one module with
$(n+1)$ commands for all $n$ states of any activity pattern and for
one (dummy) initial state.  The initial state $(s=0,k=0)$ is a dummy
that encodes the global initial distribution $\initdist$ for the user
activity patterns. All activity patterns have the same set of states
and we enumerate them from $1$ to $n$; we can label them conveniently
with atomic propositions. For instance, in a Hungry Yoshi UMM the
states $(0,k)$ to $(4,k)$ are labelled by the atomic proposition
$\init$, $\seeY$, $\feed$, $\seeP$, $\pick$ respectively.  For each
state $(s,\cdot)$, with $s>0$, we have a command defining all possible
$n\cdot K$ probabilistic transitions. If the probability of an update
is null, then the corresponding update, i.e., transition, does not
take place.
\begin{figure}[!t]
\centering
{\small
$
\begin{array}{l}
  \mathbf{module~} \mathit{UserMetamodel\_m}\\
  \quad  s:[0\,..\,n]~\mathbf{init}~0;\\
  \quad  k:[0\,..\,K]~\mathbf{init}~0;\\ \\

  \quad []~ (s=0) \lra \theta_m(1)*\iota^{\mathit{init}}(1):(s'=1)\,\&\,(k'=1) +
  \ldots + \\
  \qquad\qquad\qquad\quad\
  \theta_m(K)*\iota^{\mathit{init}}(n):(s'=n)\,\&\,(k'=K);\\ 
  \quad  []~ (s=1) \lra \theta_m(1)*\proba_1(1,1): (s'=1)\,\&\,(k'=1) + \ldots
  +\\
  \qquad\qquad\qquad\quad\ \theta_m(K)*\proba_K(1,n): (s'=n)\,\&\,(k'=K); \\
  \quad  \vdots\\
  \quad  []~ (s=n) \lra \theta_m(1)*\proba_1(n,1): (s'=1)\,\&\,(k'=1) + \ldots
  +\\
  \qquad\qquad\qquad\quad\ \theta_m(K)*\proba_K(n,n): (s'=n)\,\&\,(k'=K); \\
  \mathbf{endmodule}
\end{array}
$
}
\caption{PRISM module for encoding a user metamodel where $K$ is the
  number of activity patterns, $n$ is the number of states in each
  user activity pattern model $\alpha_k$, $\iota^{\mathit{init}}$ is
  the initial distribution in $\alpha_k$, $\proba_k(i,j)$ is the
  transition probability from state $i$ to state $j$ in $\alpha_k$,
  and $\theta_m(k)$ is the probability of user $m$ to choose the
  activity patterns $\alpha_k$, for all $i,j\in\{1, \ldots, n\}$,
  $k\in\{1,\ldots,K\}$. }
\label{fig:prism-encoding}
\end{figure}

\section{Analysing the Hungry Yoshi UMM}\label{analysis}
\def\wl{\par \vspace{\baselineskip}} 

In this section, we give some example analysis for a UMM.  Namely, we
encode and evaluate quantitatively several example questions from
Sect.~\ref{questions} for the UMM with the strategy for transitioning
between activity patterns defined by $\theta = (0.7,0.3)$. The PRISM
models and property files are freely available\footnote{Available
  from~\url{http://dcs.gla.ac.uk/~oandrei/yoshi} .}.

Recall that to score highly, a user must feed one or more yoshis (the
appropriate fruit) often. An informal inspection of $\alpha_1$ and
$\alpha_2$ indicates that $\alpha_2$ is a less effective strategy for
playing the game, since paths from $\seeP$ and $\seeY$ to $\feed$ are
unlikely.  Now, by formal inspection of the UMM (encoded in \PRISM),
we can investigate this hypothesis more rigorously.  We consider
properties that are parametrised by a number of button taps (e.g.
$N$, $N1$, $N2$) and by activity pattern (e.g.~$\alpha_1$,
$\alpha_2$), so we use the \PRISM\ experiment facility that allows us
to evaluate and then plot graphically results for a range of formulae.

\begin{figure}[!t]
  \centering
  \subfigure{\includegraphics[width=0.45\linewidth]{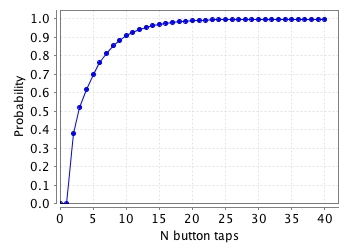}}
  \quad
  \subfigure{\includegraphics[width=0.45\linewidth]{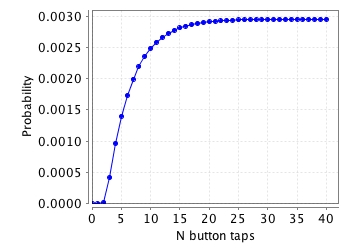}}
  \caption{Question \ref{q:feedYoshi}: the probability of feeding a
    yoshi for the first time within $N$ button taps for the activity
    pattern $\alpha_1$ on the left  and for $\alpha_2$ on the right.}
    \label{fig:feedYoshi}
\end{figure}

\refstepcounter{myquestion}\label{q:feedYoshi}
\paragraph{Question \ref{q:feedYoshi}.} How many button taps $N$ does
it take to feed a yoshi for the first time? We encode this by the
probabilistic until formula:
$$\begin{array}{rcl}
  p_{\ref{q:feedYoshi}}(i) & = & \Probs{\umm}{\init}((\lnot\feed) \,\TU^{\leq
    N}\, ((\alpha=i)\land\feed))
\end{array}
$$
For activity pattern $\alpha_1$, Figure~\ref{fig:feedYoshi} shows that
within 2 button taps the probability increases rapidly, and after 5
button taps the probability is more than $70\%$.  Contrast this with
the results for $\alpha_2$: the probability increases rapidly after 3
button taps but soon it reaches the upper bound of 0.003. Comparing
the two results, $\alpha_1$ is clearly more effective.
 
\bigskip

Now we consider more complex questions concerning sequences of feeding
and picking; recall that a basket can hold at most 5 fruits and extra
points are gained by feeding a yoshi its required 5 fruits without any
other interruption.  In Question~\ref{q:5feeds} we consider feeding a
full basket to a yoshi, without being interrupted by a $\pick$ (note,
we allow $\seeY$ and $\seeP$); in Question~\ref{q:5picks5feeds} we
consider picking a full basket, without being interrupted by a
$\feed$, followed by feeding the full basket to a yoshi, without being
interrupted by a $\pick$ (again, we allow $\seeY$ and $\seeP$).

\refstepcounter{myquestion}\label{q:5feeds}
\paragraph{Question \ref{q:5feeds}.}
We calculate the probability of reaching the state $\feed$ within $N$
button taps and then visiting it (with the same activity pattern) for
another four times without visiting the state $\pick$, for each
activity pattern ($i\in\{1,2\}$):
$$ 
\begin{array}{rcl}
  p_{\ref{q:5feeds}}(i) & = & \Probs{\umm}{\init}(\TF^{\leq N}((\alpha=i)\land
  \feed)) \cdot  (\Probs{\umm_{\mid 
      \alpha=i}}{\feed}(\TX((\lnot\pick\land \lnot\feed)
  \,\TU\,\feed))^4 
\end{array}
$$
As example, in this case we also give the corresponding PRISM formula:
{\footnotesize
\begin{verbatim}
P=?[F<=N((alpha=i)&"feed")]*
pow(filter(min,P=?[X(((alpha=i)&(!"pick")&(!"feed"))U((alpha=i)&"feed"))], 
           ((alpha=i)&"feed")),4)
\end{verbatim}
}

\begin{figure}[!ht]
  \begin{minipage}[b]{0.45\linewidth}
    \centering
    \includegraphics[width=\linewidth]{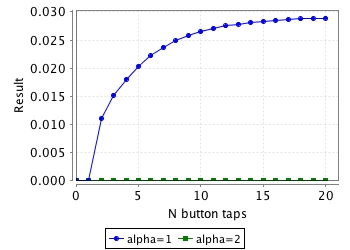}
    \caption{Question \ref{q:5feeds}: the probability of feeding a
      yoshi the whole fruit basket with no fruit picking in between.}
    \label{fig:5feeds}
  \end{minipage}
  \hspace{0.5cm}
  \begin{minipage}[b]{0.45\linewidth}
    \centering
    \includegraphics[width=\linewidth]{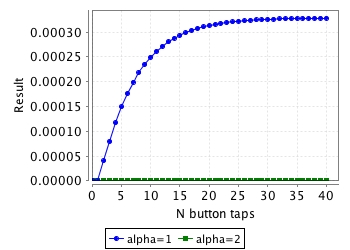}
    \caption{Question \ref{q:5picks5feeds}: the probability of picking
    five pieces of fruit and then feed a yoshi the whole basket.}
    \label{fig:5picks5feeds}
  \end{minipage}
\end{figure}

\noindent
The results are shown in Fig.~\ref{fig:5feeds} for both activity
patterns and a range of number of button taps. While the results for
$\alpha_1$ (converging to 0.03) are higher than for $\alpha_2$
(effectively 0); they are both small.  There could be several causes
for this.  For example, players are only made aware of the possibility
of extra points at the end of the instructions pages, or available
fruit depends on the external environment.  If designers/evaluators
want this investigated further, then we would require to record and
extract more detail from the logs, for example to log numbers of
available WiFi access points and scrolls through instruction pages.

\refstepcounter{myquestion}\label{q:5picks5feeds}
\paragraph{Question \ref{q:5picks5feeds}.}

We calculate the probability of reaching the state $\feed$ only after
visiting the state $\pick$ five times (without feeding) and then
visiting the state $\feed$ four more times (without picking), for each
activity pattern:
$$ 
\begin{array}{rcl}
  p_{\ref{q:5picks5feeds}}(i) & = &  \Probs{\umm}{\init}[(\lnot
  \pick) \,\TU^{\leq N}((\alpha=i) \land \pick)] \cdot \\
  & & (\Probs{\umm_{\mid \alpha=i}}{\pick}[X((\lnot\feed \land \lnot\pick)
  \,\TU\,(\pick))])^4\cdot\\ 
  && \Probs{\umm_{\mid
      \alpha=i}}{\pick}[(\lnot\feed) \,\TU\, \feed] \cdot
  (\Probs{\umm_{\mid \alpha=i}}{\feed}[X((\lnot\pick) \land (\lnot\feed))
  \,\TU\,\feed)])^4 
\end{array}
$$
The results are presented in Fig.~\ref{fig:5picks5feeds}. Again, while
the probabilities are low (presumably for the reasons outlined above
for Question~\ref{q:5feeds}) the user that picks a full basket and
feeds it to a yoshi by following activity pattern $\alpha_1$ does it
with around 0.0003 probability within 15 steps into the game, whereas
if they follow $\alpha_2$ from the beginning, they never empty the
basket.  So again, $\alpha_1$ proves to be more effective.

\smallskip

Now we turn our attention to a question that involves a {\em change}
of activity pattern, i.e. a change in the playing strategy.

\begin{figure}[!t]
  \centering
  \subfigure{\includegraphics[width=0.45\linewidth]{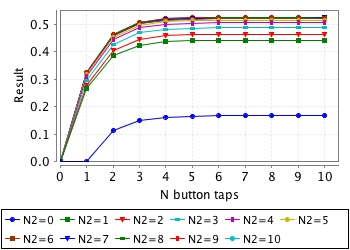}}
  \quad
  \subfigure{\includegraphics[width=0.45\linewidth]{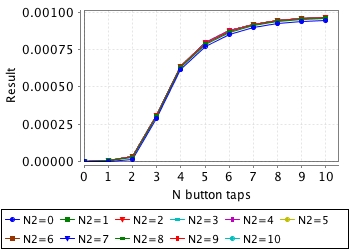}}
  \caption{Question \ref{q:changePattern} for $N_2\leq 10$ and $i=1$
    on the left and $i=2$ on the right.}
    \label{fig:changePatternBounded}
\end{figure}

\begin{figure}[!t]
  \centering
  \subfigure{\includegraphics[width=0.45\linewidth]{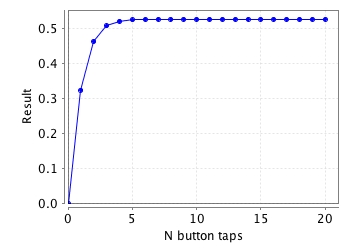}}
  \quad
  \subfigure{\includegraphics[width=0.45\linewidth]{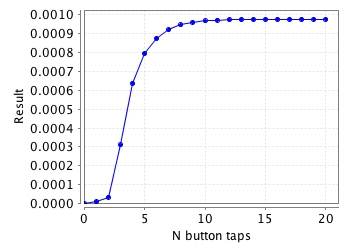}}
  \caption{Question \ref{q:changePattern} for $N_2=\infty$ and $i=1$
    on the left and $i=2$ on the right.}
    \label{fig:changePatternUnbounded}
\end{figure}

\refstepcounter{myquestion}\label{q:changePattern}
\paragraph{Question \ref{q:changePattern}.}
What is the probability of starting with an activity pattern and not
feeding a yoshi within $N$ button taps, then changing to the other
activity pattern and eventually first feeding a yoshi within $N_2$ button
taps? We compute this probability as follows, where
$\mathcal{L}_0=\{\feed, \pick, \seeY, \seeP\}$:
$$
\begin{array}{rcl}
  p_{\ref{q:changePattern}}(i) & = & \sum_{\ell\in
    \mathcal{L}_0}\Probs{\umm}{\init}((\lnot(\alpha=i) \land  
  \lnot\feed)\, \TU^{\leq N} ((\alpha=i) \land \ell)) \cdot 
\\ &&  \quad\qquad 
\Probs{\umm_{\mid \alpha=i}}{\ell}((\lnot \feed)\,\TU^{\leq N_2} \feed)
\end{array}
$$
Figure~\ref{fig:changePatternBounded} shows the results for switching
from activity patterns $\alpha_1$ to $\alpha_2$ and vice-versa
respectively for less than 10 button taps to feed a yoshi after
switching the activity pattern, while
Figure~\ref{fig:changePatternUnbounded} shows the same but for an
unbounded number of button taps (to feed a yoshi).  We can see that
success is much more likely by switching from $\alpha_2$ to
$\alpha_1$, than switching from $\alpha_1$ to $\alpha_2$, and a user
needs about 4-5 button taps to switch from $\alpha_2$ to $\alpha_1$ to
maximise their score. This latter result is not surprising,
considering that users might first inspect the game, which would
involve visiting the 4 states.

\bigskip

All analyses were performed on a standard laptop.  Note that for
brevity, the mobile app analysed here, and its formal model, are
relatively small in size; more complex applications will yield more
meaningful activity patterns and complex logic properties that can be
analysed on the metamodels. While state-space explosion of the UMM
could be an issue, it is important to note that the state-space does
not depend on the number of users, but on the granularity of the
states (logged in-app actions) we distinguish.

\section{Discussion}\label{discussion}
We reflect upon the results obtained for the Hungry Yoshi example and
further issues raised by our approach.

\paragraph{Hungry Yoshi usage.}
Our analysis has revealed some insight into how users have actually
played the game: $\alpha_{1}$ corresponds to a more successful game
playing strategy than $\alpha_2$ and a user is much more likely to be
effective if they change from $\alpha_{2}$ to $\alpha_{1}$ (rather
than vice-versa), thus we conclude that $\alpha_{1}$ is {\em expert}
behaviour and $\alpha_{2}$ is {\em ineffective} behaviour.  (Note that
users can, and do, switch between both behaviours, e.g.  a user who
exhibits expert behaviour can still exhibit ineffective behaviour at
some later time.) This interpretation of activity patterns can inform
a future redesign that helps users move from ineffective to expert
behaviour, or induces explicitly populations of users to follow
selected computation paths to reach certain goal states.  We note that
the developers had very little intuition about how often, or if, users
were picking a full basket and then feeding a yoshi (e.g.
Questions~\ref{q:5picks5feeds} and~\ref{q:changePattern} in
Sect.~\ref{analysis}), and so the results, which indicate this
scenario is quite rare, provided a new and useful insight for them.

\paragraph{Temporal properties.}
The properties refer to propositions about user-initiated events
(e.g.~{\em seeY, feed}) and activity patterns (e.g.~$\alpha_{1}$,
$\alpha_{2}$). A future improvement would be a syntax that
parametrises the temporal operators by activity pattern. We note that
PCTL properties alone were insufficient for our analysis and we have
made extensive use of {\em filtered} properties. We also note that for
some properties we have used \PRISM\ rewards, e.g. to compare scores
between activity patterns, but these are omitted in this short paper.

\paragraph{Reasoning about users.}
Model checking is performed on the UMM resulting from the augmentation
of the set of $K$ activity patterns with a strategy $\theta_{m}$.  It
is simple to select a user by selecting a $\theta_{m}$ and to analyse
the resulting UMM.  Metrics on the set $\{\theta_{m} \mid m=1,\ldots,M\}$
will be used in future work to characterise how the results of the
analysis change depending on the value of one $\theta_{m}$, in the
hope that results of the analysis for one user can be generalised to
users close by (under the given metric).

\paragraph{Formulating hypotheses: domain specific and generic.}
We have considered domain-specific hypotheses presented by developers
and evaluators, but could a formal approach help with hypothesis
generation?  For example, we could frame questions using the
specifications patterns for probabilistic quality properties as
defined in~\cite{Grunske08} (probabilistic response, probabilistic
precedence, etc.).  Referring to our questions in
Sect.~\ref{questions}, we recognise in the first item the {\em
probabilistic precedence} pattern, in the second one the {\em
probabilistic response} pattern, and in the last two the {\em
probabilistic constrained response} pattern.  However, these patterns
refer only to the top level structure, whereas all our properties
consist of multiple levels of embedded patterns. Perhaps more complex
patterns are required for our domain?  The patterns
of~\cite{Grunske08} were abstracted from avionic, defence, and
automotive systems, which are typical reactive systems; does the
mobile app domain, or domains with strong user interaction exhibit
different requirements?  We remark also that analysis of activity
patterns is just one dimension to consider: there are many others that
are relevant to tailoring software to users, for example software
variability and configuration, and user engagement.  These are all
topics of further work.

\paragraph{Choosing $K$ activity patterns.}
What is the most appropriate value for $K$, can we guide its choice?
While we could use model selection or non-parametric methods to infer
it, there might be domain-based reasons for fixing $K$.  For example,
we can start with an estimate value of $K$ and then compare analysis
activity patterns: if properties for two different activity patterns
give very close quantitative results then we only need a smaller $K$.

\paragraph{What to log?}
This is a key question and depends upon the propositions we examine in
our properties, as well as the overheads of logging (e.g. on system
performance, battery, etc.) and ethical considerations (e.g. what have
users agreed). Formal analysis will often lead to new instrumentation
requirements, which in turn will stimulate new analysis.  For example,
our analysis of Hungry Yoshi has indicated a need for logged traces to
include more information about current context, e.g. the observable
access points (yoshis).

\paragraph{Overall approach and future plans.}
In this paper we have focused on defining the appropriate statistical
and formal models, their encoding, and reasoning using model checking.
We have not explored here the types of insights we can gain into user
behaviours from our approach, nor how we can employ these in system
redesign and future system design, especially for specific
subpopulations. Further, in this short paper, we have not considered
the role of prediction from analysis and the possibilities afforded by
longitudinal analysis. For example, how do the activity patterns and
properties compare between users in 2009 and users in 2013?  This is
ongoing work within the {\em A Population Approach to Ubicomp System
Design} project, where we are working with system developers on the
practical application of our formal analysis in the design and
redesign of several new apps.  We are also investigating metrics of
user engagement, tool support, and integration of this work with
statistical and visualisation tools.

\section{Related work} \label{related}

Our work is a contribution to the new {\em software analytics}
described in~\cite{MenziesZ13}, focusing on local methods and models,
and user perspectives. It is also resonates with their prediction that
by 2020 there will be more use of analytics for mobile apps and
games. Recent work in analysis of user behaviours in systems,
especially XBox games, is focused on understanding how features are
used and how to drive users to use desirable features.  For example,
\cite{huang-chi-2013} investigates video game skills development for
over 3 million users based on analysis of users' TrueSkill
rating~\cite{trueskill}.  Their statistical analysis is based on a
single, abstract ``skill score'', whereas our approach is based on
reasoning about computation paths relating to in-app events and
temporal property analysis of activity patterns.  Our approach can be
considered a form of run-time quantitative verification (by
probabilistic model checking) as advocated by Calinescu et
al. in~\cite{CalinescuGKM12}.  Whereas they consider functional
behaviour of service-based systems (e.g. power management) and
software evolution triggered by a violation of correctness criteria
because software does not meet the specification, or environment
change, we address evolution based on behaviours users {\em actually}
exhibit and how these behaviours relate to system requirements, which
may include subtle aspects such as user goals and quality of
experience.  Perhaps of more relevance is the work on off-line runtime
verification of logs in~\cite{StollerBSGHSZ11} that estimates the
probability of a temporal property being satisfied by program
executions (e.g. user traces).  Their approach and results could help
us determine how logging sampling in-app actions and app configuration
affects analysis of user behaviour. Finally we note the very recent
work of~\cite{GhezziPST14} on a similar approach and comment the major
differences in Sect.~\ref{inference}. In addition they analyse REST
architectures (each log entry corresponds to a web page access),
whereas the mobile apps we are analysing are not RESTful, we can
include more fine grained and contextual data in the logged user data.

\section{Conclusions and Future Work}\label{conclusions}

We have outlined our contribution to software analytics for user
interactive systems: a novel approach to probabilistic modelling and
reasoning over actual user behaviours, based on systematic and
automated logging and reasoning about users.  Logged user traces are
computation paths from which we infer {\em activity patterns},
represented each by a discrete-time Markov chain (DTMC). A user meta
model is deduced for each user, which represents users as mixtures
over DTMCs.  We encode the user metamodels in the probabilistic model
checker \PRISM\ and reason about the metamodel using probabilistic
temporal logic properties to express hypotheses about user behaviours
and relationships within and between the activity patterns.
State-space size is independent of the size of population of users, and  is 
related only to the number of distinct events that are logged.

We motivated and illustrated our approach by application to the Hungry
Yoshi mobile iPhone game, which has involved several 
thousands of users worldwide.  We showed how to encode some example
questions posed by developers and evaluators in a probabilistic
temporal logic, and obtained quantitative results for an example user
metamodel.  After considering our formal analysis of two activity
patterns, we conclude the two activity
patterns distinguish {\em expert} behaviour from {\em ineffective} behaviour
and represent different strategies about how to play the
game.  While in this example the individual activity pattern DTMCs are
small in number and size, in more complex settings it will be
impossible to gain insight into behaviours informally, and in
particular to insights into relationships between the activity
patterns, so automated formal analysis of the UMM will be essential.

Further work will focus on gaining experience of practical application
of our approach within redesign, and tying in with the other analytic
methods we have at hand, for example visualisation.  We will also
combine the orthogonal concerns of user activity patterns and software
configuration and structural variability.  We will consider software
configurations in use and distributions of software {\em features} (a
feature~\cite{CalderKMR03,ClassenHS08} is a component that adds new
functionality).  Properties to examine include: Which configurations
and contexts are most popular for which types of users?  Which are
problematic? How are some particular configurations used and do they
lead to longer or better user engagement?  How does my setup compare
to my friends? Are there configurations like my own that don't crash
so much and how are they used?

\vspace{.3cm}
 \noindent
 {\bf Acknowledgments.} This research is supported by EPSRC Programme
 Grant {\em A Population Approach to Ubicomp System Design}
 (EP/J007617/1). The authors thank all members of the project, and
 Gethin Norman for fruitful discussions.

\bibliographystyle{splncs}

\end{document}